\documentclass[3p,times,twocolumn]{elsarticle}

\usepackage{lineno}
\usepackage{xcolor}   
\usepackage{graphicx}
\usepackage{ulem}  

\def\um{\ifmmode {\mathrm{\mu m}}\else
                  \textrm{$\mu$m }\fi}%
\def\GeV{\ifmmode {\mathrm{\ Ge\kern -0.1em V}}\else
                   \textrm{Ge\kern -0.1em V}\fi}%
\def\MeV{\ifmmode {\mathrm{\ Me\kern -0.1em V}}\else
                   \textrm{Me\kern -0.1em V}\fi}%
\def\keV{\ifmmode {\mathrm{\ keg\kern -0.1em V}}\else
                   \textrm{ke\kern -0.1em V}\fi}%
\def\eV{\ifmmode  {\mathrm{\ e\kern -0.1em V}}\else
                   \textrm{e\kern -0.1em V}\fi}%

\def\uW{\ifmmode  {\mathrm{\mu  W}}\else
                   \textrm{$\mu$W}\fi}%

\begin{document}
\begin{frontmatter}

\title {
Optical transceivers for event triggers in the ATLAS phase-I upgrade
  }

\author[ccnu,smu]{L. Zhang}
\author[smu]{C. Chen}
\author[smu]{I. Cohen}
\author[smu]{E. Cruda}
\author[smu]{D. Gong}
\author[ipas]{S. Hou}
\author[umich]{X. Hu}
\author[ccnu,smu]{X. Huang}
\author[ipas,ntu]{J.-H. Li}
\author[smu]{C. Liu}
\author[smu]{T.~Liu}
\author[smu]{L.~Murphy}
\author[umich]{T.~Schwarz} 
\author[ccnu,smu]{H. Sun}
\author[ccnu]{X. Sun}
\author[smu]{J. Thomas}
\author[smu]{Z. Wang}
\author[smu]{J. Ye}
\author[ccnu,smu]{W. Zhang}


\address[ccnu]{  Central China Normal University, Wuhan, Hubei 430079, China }
\address[smu]{   Southern Methodist University, Dallas, TX 75275 USA }
\address[ipas]{  Academia Sinica, Taipei, Taiwan 11529 }
\address[umich]{ University of Michigan, Ann Arber, MI 48109  USA  }
\address[ntu]{   National Taiwan University, Taipei, Taiwan 10617 }

\begin{abstract}

The ATLAS phase-I upgrade aims to enhance event trigger performance in 
the Liquid Argon (LAr) calorimeter and the forward muon spectrometer. 
The trigger signals are 
transmitted by optical transceivers at 5.12~Gbps per channel 
in a radiation field.
We report the design, quality control in production and ageing test of the
transceivers fabricated with the LOCld laser driver and multi-mode 
850 nm vertical-cavity surface-emitting laser (VCSEL).
The modules are packaged in miniature formats of dual-channel 
transmitter (MTx) and transceiver (MTRx) for the LAr. 
The transmitters are also packaged in small form-factor pluggable (SFP)
for the muon spectrometer.  
In production, the LOCld chips and VCSELs in TOSA package
were examined before assembly.
All of the modules were tested and selected during production for
quality control based on the eye-diagram parameters {\color{black} of outputs}. 
The yield is 98 \% for both the MTx and MTRx 
{\color{black} on} a total 4.7k modules. 
The uniformity of transmitter channels of a MTx 
{\color{black} was assured by choosing the TOSA components with} 
approximately equal light powers. 
The ageing effect is monitored in burn-in of a small batch of 
transmitter modules with bit-error test and eye-diagrams measured 
periodically. The observables are stable with the light power
degradation within 5 \% over a period of more than 6k~hours.

\end{abstract}
\end{frontmatter}

\section{Introduction}

The ATLAS Phase-I upgrade aims to enhance event trigger performance \cite{P1}
in the Liquid Argon (LAr) calorimeter \cite{LAr} and the forward muon 
New Small Wheel spectrometer (NSW) \cite{NSW}.
The {\color{black} trigger} 
signals are transmitted by customized optical transceivers at 5.12~Gbps. 
These devices include the dual-channel miniature
optical transmitters (MTx) and transceivers (MTRx) \cite{MTx,MTRx}. 
The LAr upgrade requires (including spares) 3240 MTx  
and 810 MTRx modules for data transmission and clock/control signals,
respectively, being installed on 150 LAr Trigger Digitizer Boards (LTDBs) \cite{LTDB}. 
The NSW requires 600 MTx's (including spares) on the 256 trigger 
router boards \cite{Router}. 

The MTx and MTRx modules are developed with multi-mode 850 nm VCSEL 
(vertical-cavity surface-emitting laser) 
for data transmission over a distance of a few hundred meters.
The packaging of VCSEL and photo-detector in
TOSA/ROSA (Transmitter/Receiver Optical Sub Assembly) formats are chosen 
for light coupling to fiber-optic cables with LC connectors. 

These modules are  required {\color{black} to operate for} more than ten years
in a harsh radiation environment. 
All of the opto-electronics have been studied for radiation hardness 
{\color{black} \cite{OptoRadhard,Ageing}}.
The laser driver employed for the transmitter channel is the LOCld \cite{LOCld},
which is a custom-developed ASIC fabricated in 0.25 $\mu$m Silicon-on-Sapphire 
(SoS) CMOS process.
It is packaged in QFN-40 format for assembly.
The type of ROSA for the receiver channel on a MTRx module is a 
customized package with the photo-diode current collected by a
CERN developed GBTIA optical receiver \cite{GBTIA}.

In the following, we discuss the module design and production process
of the MTx and MTRx.
In Sec.~\ref{sec:MTx} the device assemblies are described.
The quality assurance of components including the TOSAs and LOCld chips
is discussed in Sec.~\ref{sec:QA}.
In production, all modules were examined for eye-diagrams at 
{\color{black} 5.12~Gbps for the LAr,
and 10~Gbps for the NSW, respectively.}
The quality control on modules is discussed in Sec.~\ref{sec:QC}.
The uniformity of transmitter channels with large deviation on light power
is analyzed. Results are reported in Sec.~\ref{sec:uniformity}.
A small batch of MTx's is monitored in ageing test.
The observables are presented in Sec.~\ref{sec:ageing} 
for an accumulated period of over 6k hours.
A short summary is given in Sec.~\ref{sec:sum}.

\begin{figure}[b!]    
  \vspace{-.2cm}
  \centering\includegraphics[width=.82\linewidth]{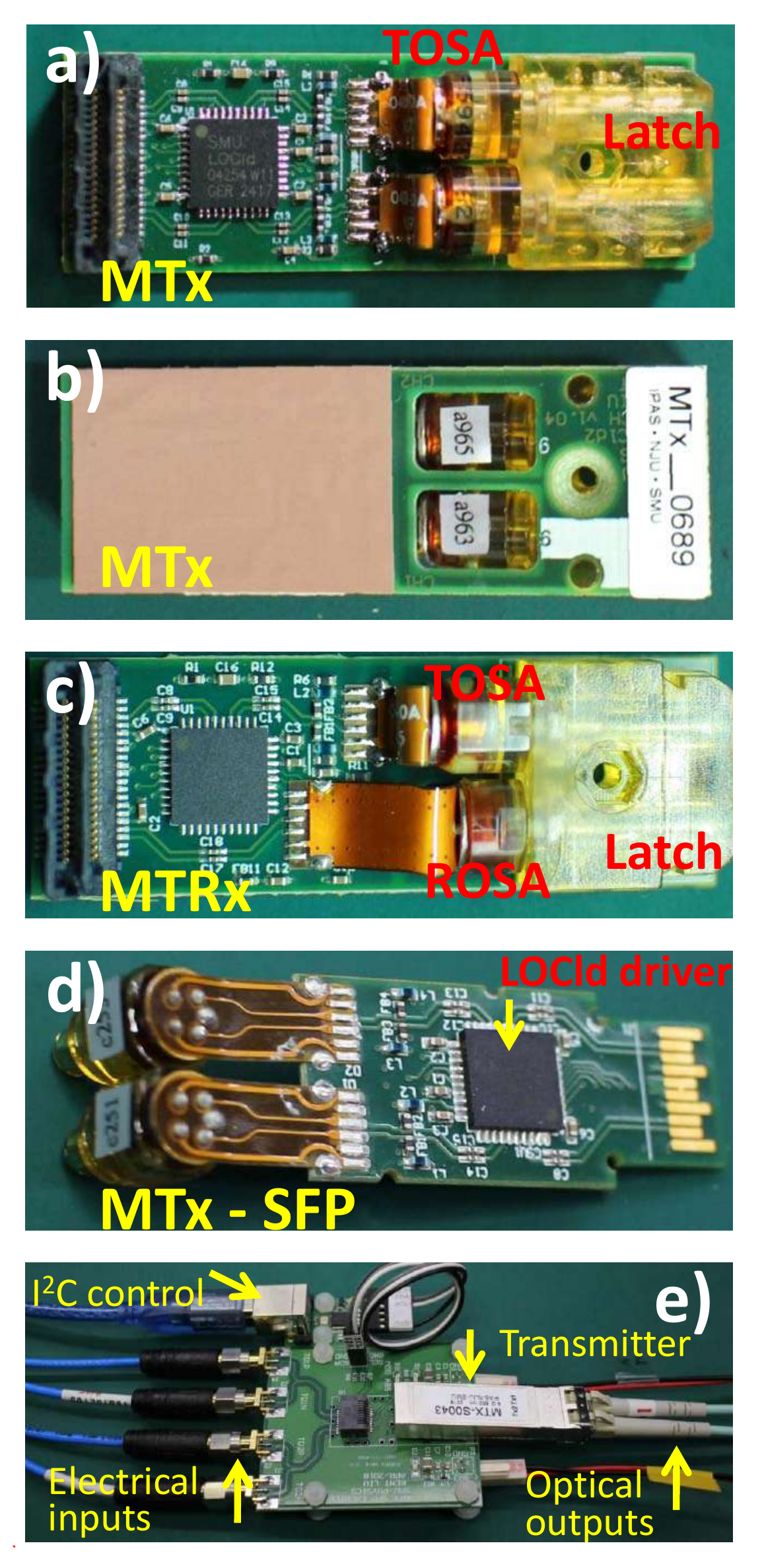}
  
  \vspace{-.2cm}
  \caption{ The MTx and MTRx modules 
  {\color{black} of lateral dimension of 
  $15 \,\mathrm{mm} \times 44 \,\mathrm{mm}$}
  are packaged with a high-density connector 
  (LSHM-120-02.5-L-DV-A-N-TR, SAMTEC)
  and a customized LC latch to a total stack height of 6 mm. 
  {\color{black} The pictures are shown for a) the front side of a MTx, 
  and b) the back side adhered with a thermal conductive pad 
  (H48-2K, t-Global). The front side of a MTRx is shown in c). 
  The assembly of a MTx in SFP format is shown in d).
  The test setup of a SFP type MTx mounted on a carrier board is 
  illustrated in e).}
  The carrier board provides connection to differential inputs and 
  I$^2$C interface to an USB module for LabVIEW control on a PC.  
  \label{fig:MTx_assembly} }
\end{figure}

\begin{figure}[t!]  
  \centering\includegraphics[width=.95\linewidth]{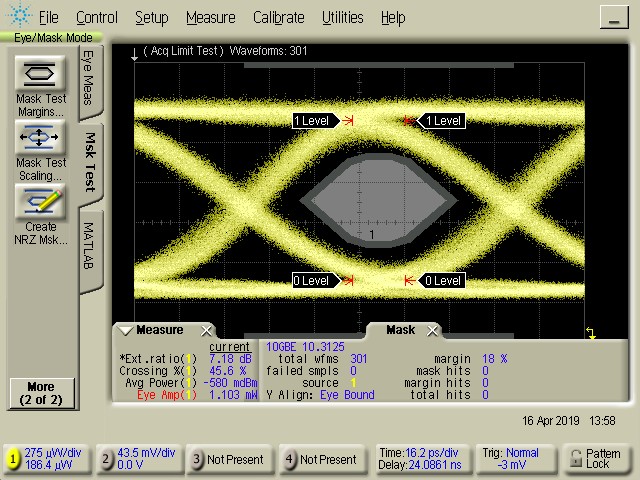}

  \caption{ {\color{black} The 10 Gbps eye-diagram is shown for a typical }
  MTx channel (measured by
  an Agilent DCA-J 86100C oscilloscope).
  The average optical power is 870~$\mu$W with the amplitude of 1.1~mW. 
  The margin to 10 Gbps mask is 18~\%. 
  \label{fig:eye10G} }
\end{figure}

\section{The MTx and MTRx modules  }
\label{sec:MTx}

The MTx and MTRx are designed for the 
restricted height of 6 mm on the LAr trigger digitizer boards. 
These modules have the same mechanical assemblies, which 
are shown in Fig.~\ref{fig:MTx_assembly}.
The pair of TOSAs on a MTx module, and the TOSA and ROSA on a MTRx 
are held by a latch for plug-in of fiber-optic cables terminated
with LC-type ferrules.
The MTx modules on the NSW router boards are mounted surrounding 
the detector in open space.  
The assembly is made in metallic SFP (small form-factor pluggable) package
for the convenience of mechanical strength and easier plug-in of
fiber-optic cables.
The MTx in SFP format 
{\color{black} and the test carrier board are}
also shown in Fig.~\ref{fig:MTx_assembly}.

The LOCld laser driver is designed with an I$^2$C interface for 
configuration of the VCSEL bias current, modulation, and peaking of light output.
The configuration is set uniformly for all modules in tests,
with the bias current to VCSEL of around 6.5 mA.
The type of TOSA employed (TTR-1F59 of the Truelight Corp.) 
has a light output specification of $0.54 - 1.02$~mW  at 6~mA.

The circuitry of modules is optimized for speed performance,
with the PCB made of FR-4 material and the passive components 
specified for 10~Gbps.  
The transmitter channels are {\color{black} evaluated} for 10~Gbps data transmission. 
Shown in Fig.~\ref{fig:eye10G} is the eye-diagram of a typical 
transmitter output. The mask margin observed is 18~\%.

\section{Quality assurance of components } 
\label{sec:QA}
 
The transmitter speed required for the ATLAS Phase-I applications is 5.12~Gbps.
The fabrication procedure has imposed selection criteria on components, in particular 
for the TOSAs and the LOCld chips.

\begin{figure}[t!]  
  \centering\includegraphics[width=.7\linewidth]{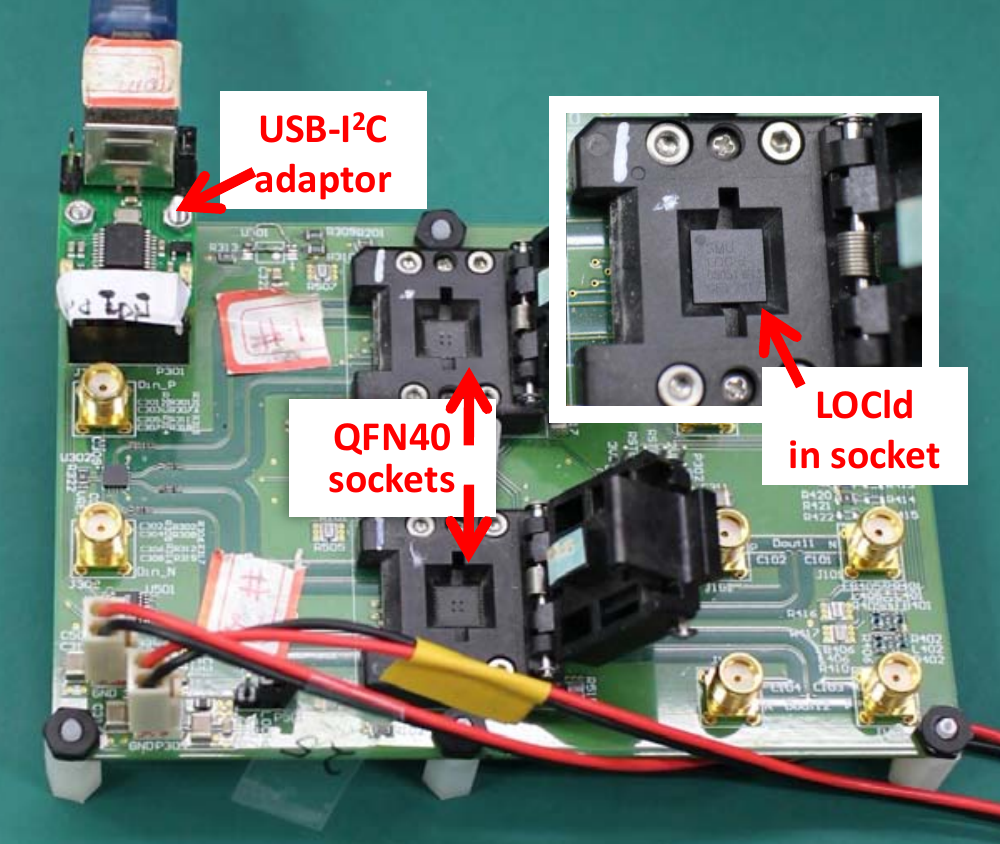}

  \caption{ \color{black} The QFN-40 packaged LOCld chips are tested in 
  matching sockets on a carrier board, which provides  
  3.3 V and 2.5 V DC powers and an USB-to-I$^2$C adapter connected to a PC.
  The LOCld configuration is conducted with a LabVIEW program. 
  \label{fig:QFNtest} }
\end{figure}

Each QFN-40 packaged LOCld chip is tested in a matching socket 
with contacts to the electrical pads of the chip. 
{\color{black} The test kit is shown in Fig.~\ref{fig:QFNtest}.}
The currents of two power supply voltages, 2.5~V and 3.3~V, are measured.
{\color{black} The VCSELs to be driven are biased from the 3.3~V,} 
and the modulations configured by the I$^2$C interface of the chip.
{\color{black}
The I$^2$C read/write  is {\color{black} conducted} 
with an USB-to-I$^2$C adapter connected to a PC.
With the total of 7200 chips tested, 26 \% had failed
due to damages in CMOS fabrication or chip packaging process. }

\begin{figure}[b!]  
  \centering
  \includegraphics[width=.49\linewidth]{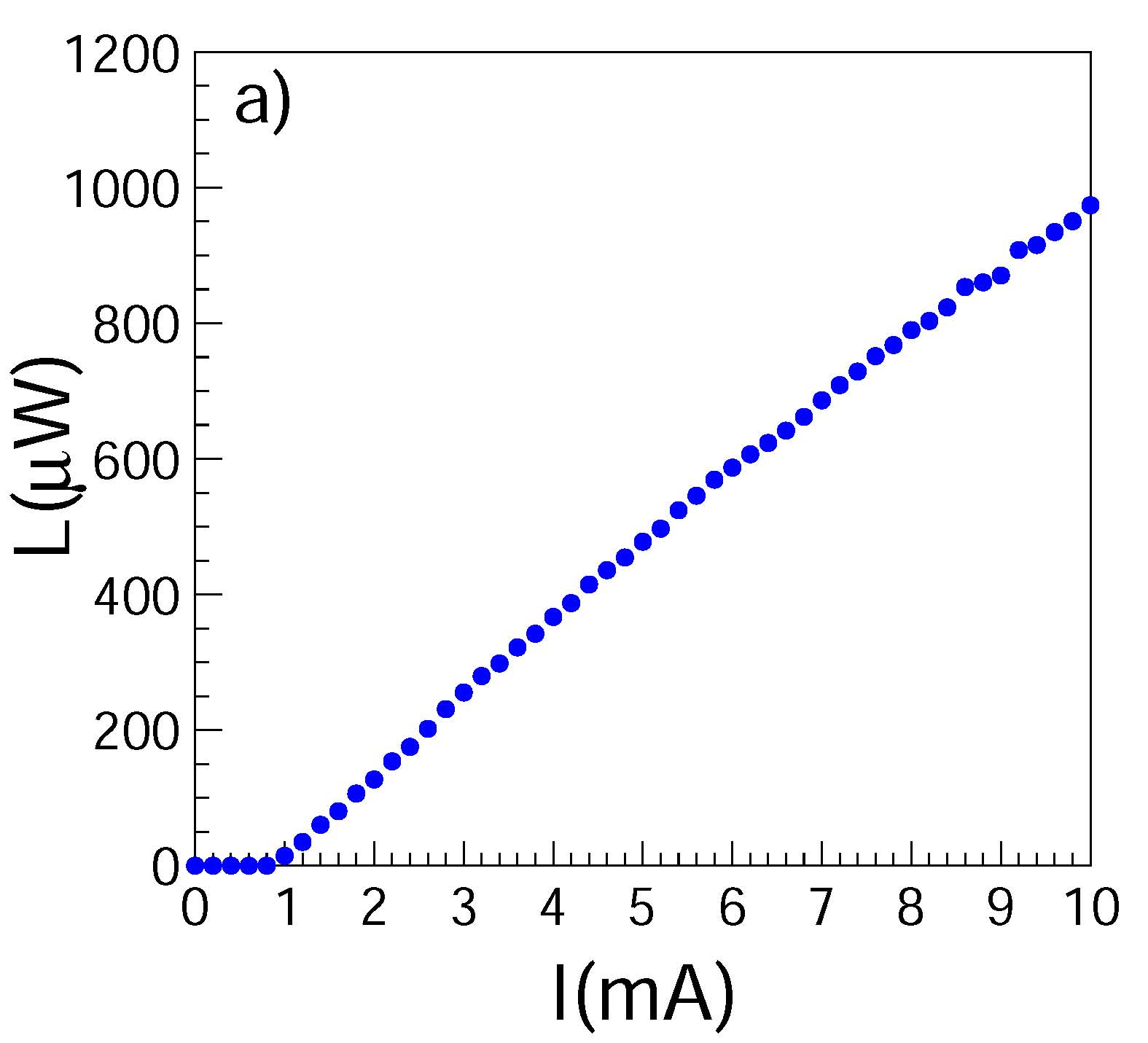}
  \includegraphics[width=.49\linewidth]{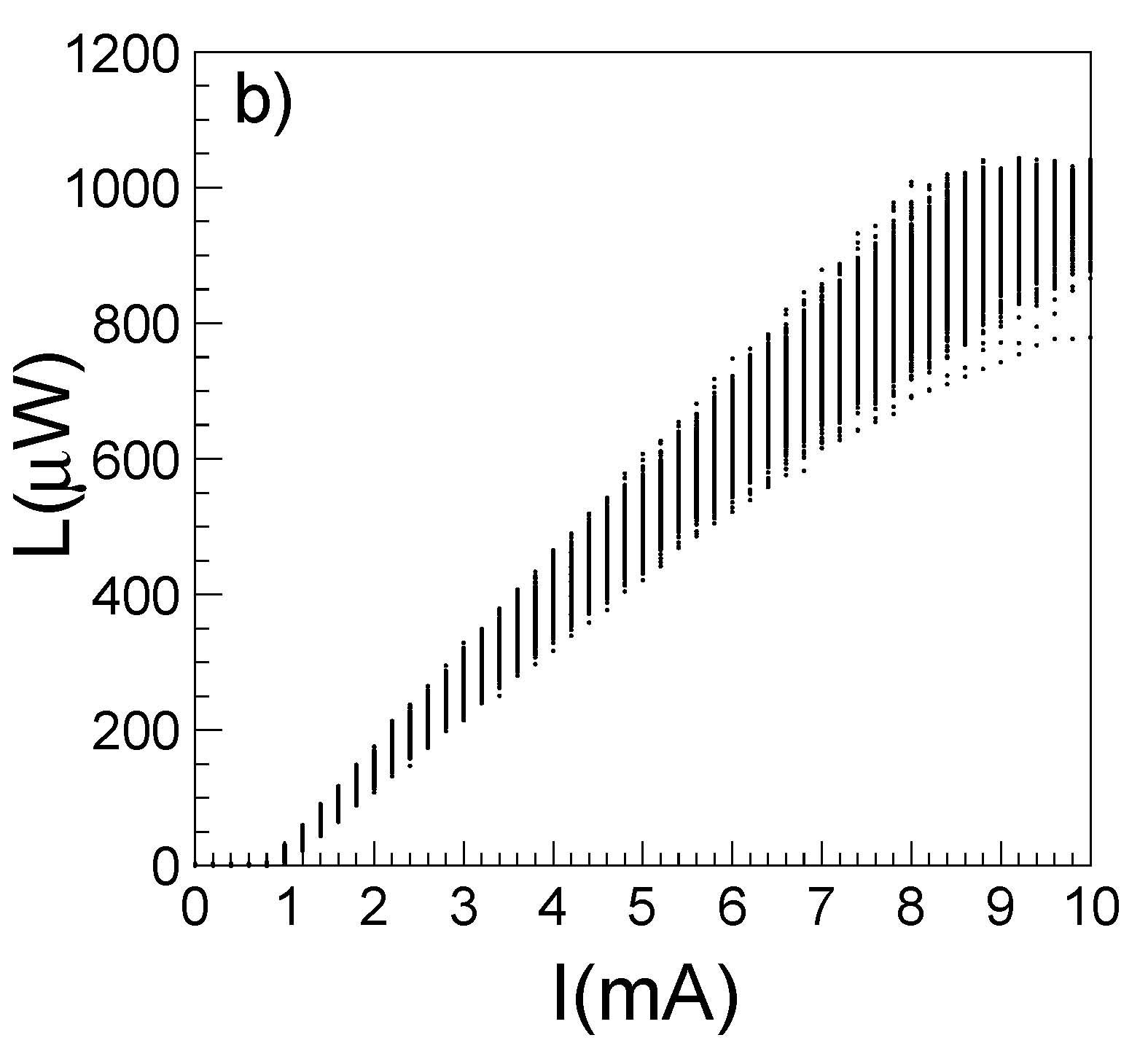} 

  \vspace{-3mm}
  \caption{ 
  {\color{black} The TOSA light output is measured with a 
  multi-mode fiber connected to a calibrated photo detector. 
  The light power versus current is shown for a) a typical channel,
  and b) the scatter plot of a batch of one thousand TOSAs.
  }
  \label{fig:TOSA-LI} }
\end{figure}

The TOSAs acquired for assembly were examined for light output versus 
current {\color{black} (L-I characteristics).}
{\color{black} The TOSA light power is measured with a multi-mode fiber 
connected to a photo detector.}
{\color{black}
The L-I distributions  are shown in 
Fig.~\ref{fig:TOSA-LI}.}
The threshold currents of VCSELs are distributed narrowly around 1.2~mA. 
The large deviation on slope efficiency 
{\color{black} is caused by {\color{black} offset} in alignment of 
the VCSEL to the TOSA lens and the systematics {\color{black} on joining}
the fiber connector to the TOSA}. 
{\color{black}The TOSAs used for module assembly are chosen
for the} light power between 550~$\mu$W and 800~$\mu$W at 6~mA.  

The fabrication of MTx and MTRx modules proceeded with pioneer runs 
of a few hundred pieces, prior to the mass production. 
The module PCBs assembled with LOCld were examined before 
TOSAs/ROSAs were soldered on, 
for the currents of the LOCld and the I$^2$C configuration.
The two TOSAs on a MTx were paired for approximately equal light powers
{\color{black} within 3 \% at 6~mA.}

The fully assembled modules were examined with each transmitter
channels measured for the light power level, 
{\color{black} optical eye-diagram,
and bit-error-rate (BER) at 10~Gbps of below} $10^{-12}$. 
A very small fraction ($<$~0.5~\%) of the transmitter
channels failed on BER, 
due to {\color{black} errors} in the LOCld data links or outputs.
{\color{black} {\color{black} Such defects} 
could not be detected till after the TOSAs were mounted.}


\begin{figure}[b!]   

  \centering\includegraphics[width=.92\linewidth]{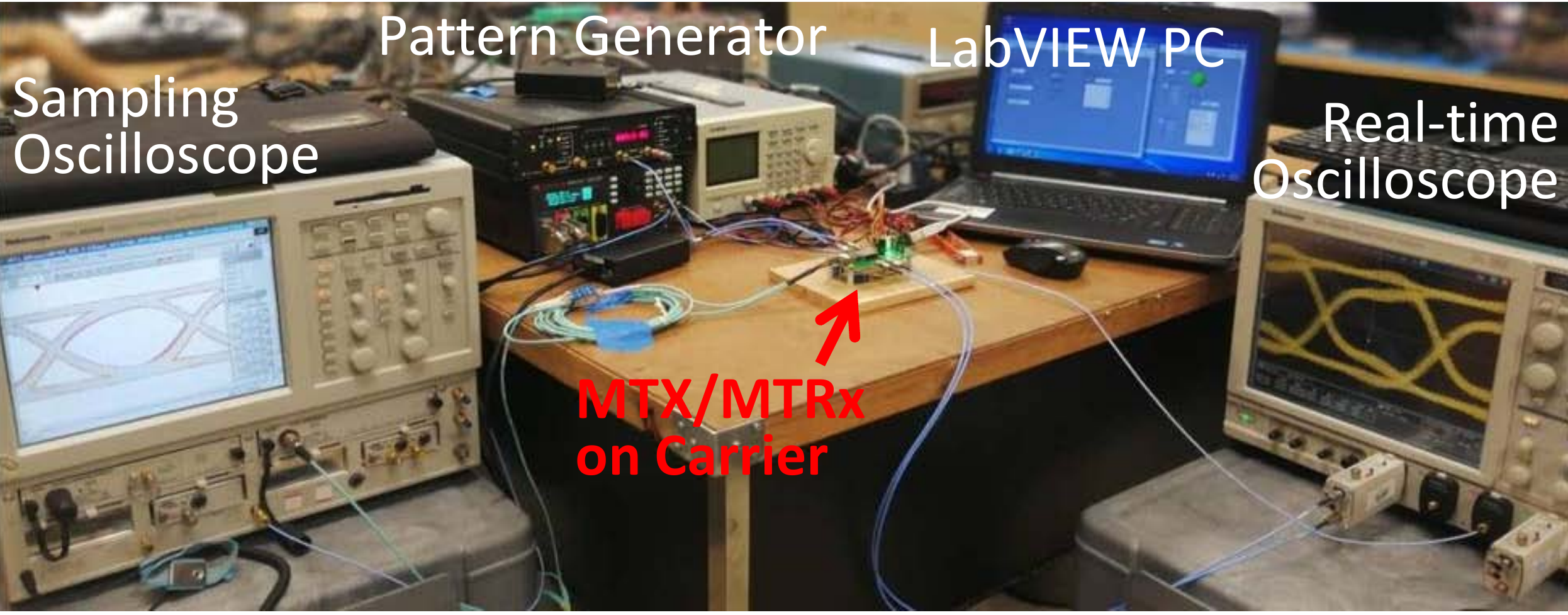}
  
  \vspace{2mm}
  \centering\includegraphics[width=.92\linewidth]{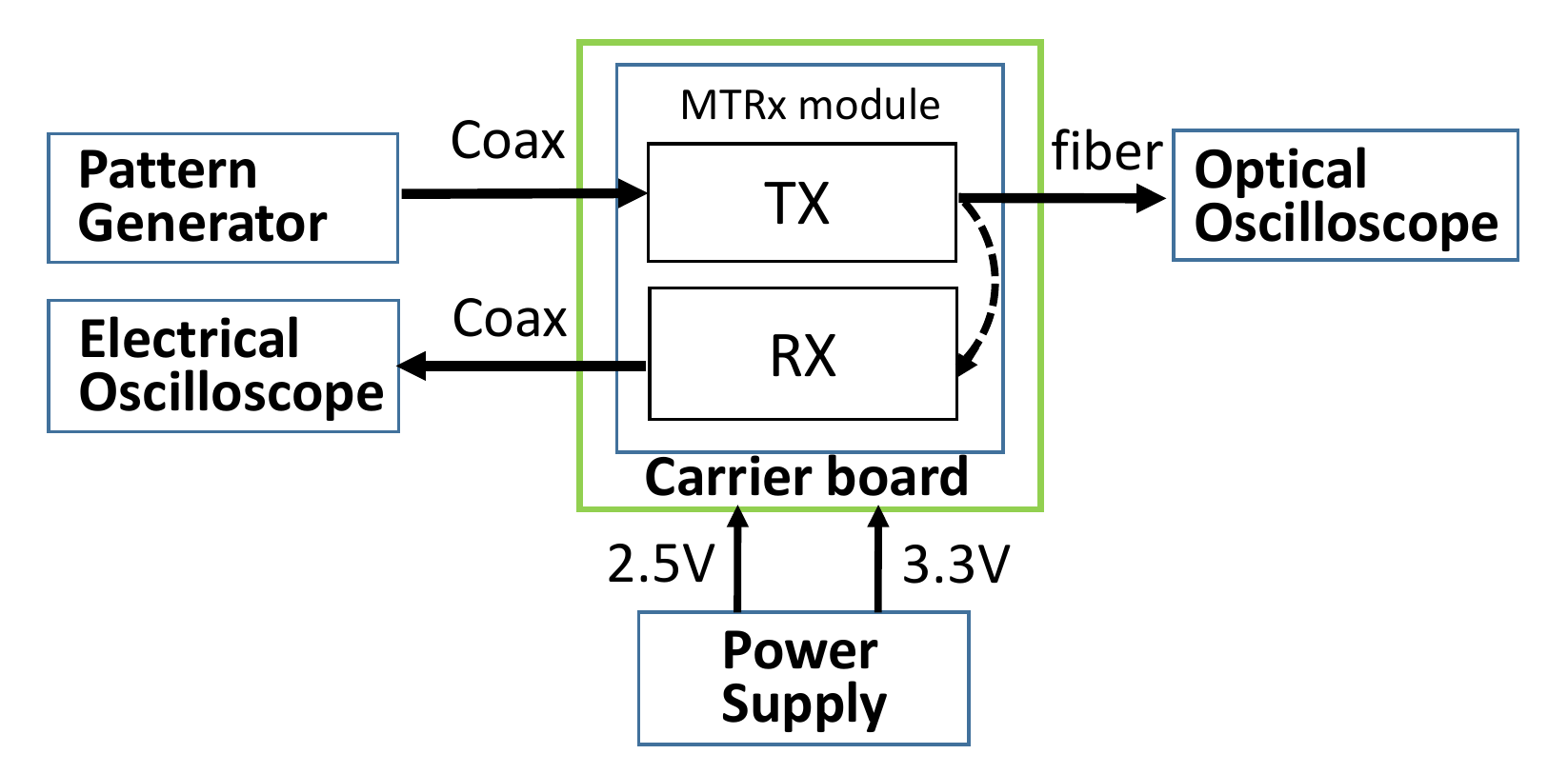}
  \caption{ The test bench is shown for quality control measuring 
  transmitter optical outputs and receiver signals of 
  MTx and MTRx modules.
  {\color{black}The setup schematics is also shown.}
  \label{fig:bench} }
\end{figure}

\begin{figure}[b!]  
  \centering\includegraphics[width=.87\linewidth]{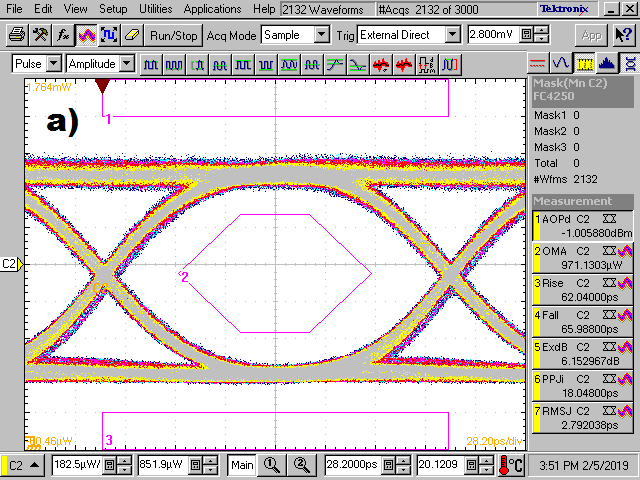}
  
  \vspace{.2cm}
  \centering\includegraphics[width=.87\linewidth]{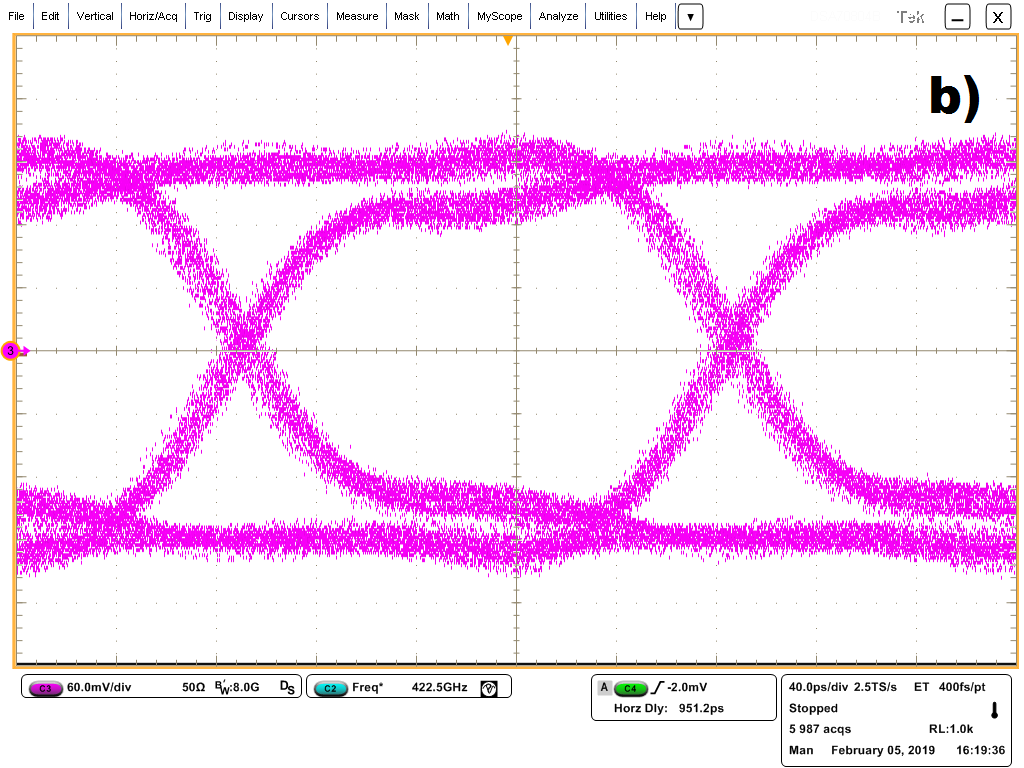}
  
  \caption{ Eye-diagrams are shown for a) the transmitter optical output, and
  b) the receiver electrical output of a typical MTRx module measured at 5.12~Gbps.
  \label{fig:QC_eyes} }
\end{figure}

\section{Quality control of modules} 
\label{sec:QC}

The MTx/MTRx modules being implemented on the LAr trigger boards 
are located inside the detector not easily accessible.
{\color{black} 
The quality control requires each module be examined   
for data transmission at the operation speed of 5.12~Gbps.}
{\color{black} The bench test setup for quality control 
is shown in Fig.~\ref{fig:bench}.}

The test conducted to a transmitter channel is configured with input
of a pseudo-random bit sequence of 2$^7 -1$  word length
from a pattern generator (PCB12500, Centellax).
The optical output is examined with the eye-diagram measured by 
a sampling oscilloscope (TDS8000B with 80C08C, Tektronix).  
For the receiver channel of a MTRx, the optical signal from the transmitter 
channel is looped back to the receiver input. 
The electrical eye-diagram is examined with a real-time oscilloscope 
(DSA72004B with a differential probe P7380SMA, Tektronix). 

The parameters of eye-diagrams are measured for the  
Average Optical Power (AOP) and Optical Modulation Amplitudes (OMA) 
of transmitter outputs, and the EXtinction ratio in decibel (EXdB), 
Root-Mean-Square Jitter (RMSJ), Rise time and Fall time of all channels. 
These parameters are recorded with a LabVIEW program on PC. 
The optical transmitter and electrical receiver eye-diagrams 
measured from a MTRx module are shown in Fig.~\ref{fig:QC_eyes}.


\begin{figure}[t!]  
  \centering\includegraphics[width=1.\linewidth]{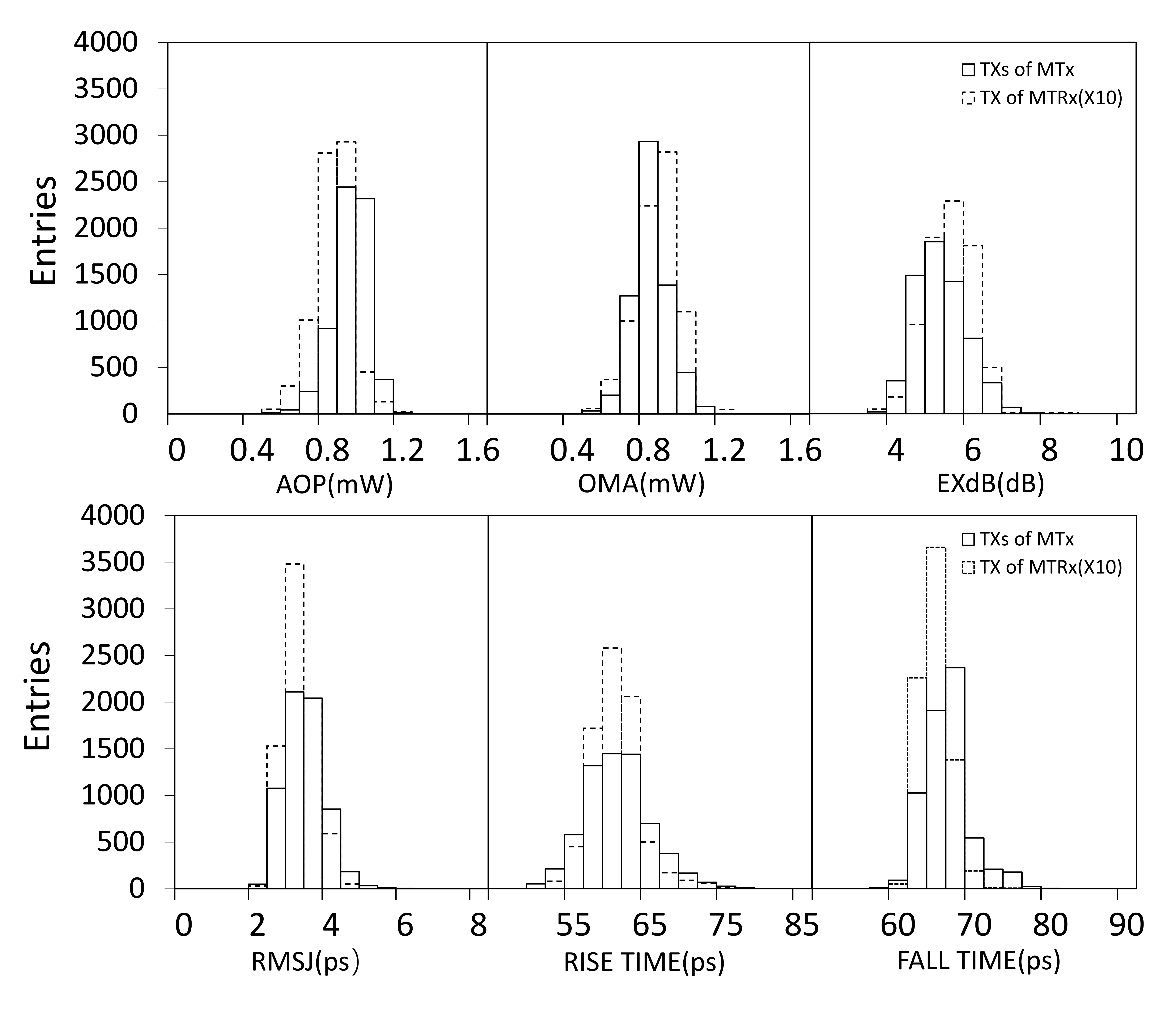}
  
  \vspace{-.5cm}
  \caption{ The distributions of eye-diagram parameters are plotted for
  the MTx and MTRx modules 
  {\color{black} for LAr} measured at 5.12~Gbps.
  \label{fig:QC} }
\end{figure}

\begin{table}[t!]
  \centering

    \begin{tabular}{c|c}  
    \hline 
       \multicolumn{2}{c}{MTRx/MTx Quality Control} \\
    \hline
       Criterion & Cut \\
    \hline
       {AOP} &  $>450$ $\mu$W  \\
       {OMA} &  $>300$ $\mu$W \\
        EXdB   &   $>3$ dB   \\
       {RMSJ} &  $<4.5$ ps  \\
       {Rise time} & $<80$ ps \\
       {Fall time} &   $<80$ ps \\
    \hline
    \end{tabular}

  \caption{ The list shows the Quality Control criteria on the eye-diagram 
   parameters of
   AOP (	Average Optical Power), 
   OMA (Optical Modulation Amplitude), 
   EXdB  (Extinction ratio),
   RMSJ (jitter RMS)  and the Rise and Fall times.
  \label{tab:QC}  }
\end{table}

{\color{black}
The selection of MTx/MTRx modules for the LAr is set on the
parameters of optical eye-diagrams with the criteria listed in Table~\ref{tab:QC}.
{\color{black} Their distributions are plotted in Fig.~\ref{fig:QC}}. }
The selection yields are 98.0~\% and 98.4~\% for the MTx and MTRx, respectively,
with the qualified modules of more than 3240 MTx and 810 MTRx produced. 

{\color{black}
The SFP type MTx modules for the NSW are selected with
each channel examined for 10~Gbps eye-diagram and light power.
Modules with any of the AOPs of the two transmitter channels 
below 550~$\mu W$, 
or the ratio of them deviated larger than 10~\% are excluded. 
The yield is 96.5~\% for the total of 600 qualified modules. 
}

\section{Uniformity of modules } 
\label{sec:uniformity}

\begin{figure}[t!] 
  \centering
  \includegraphics[width=.49\linewidth]{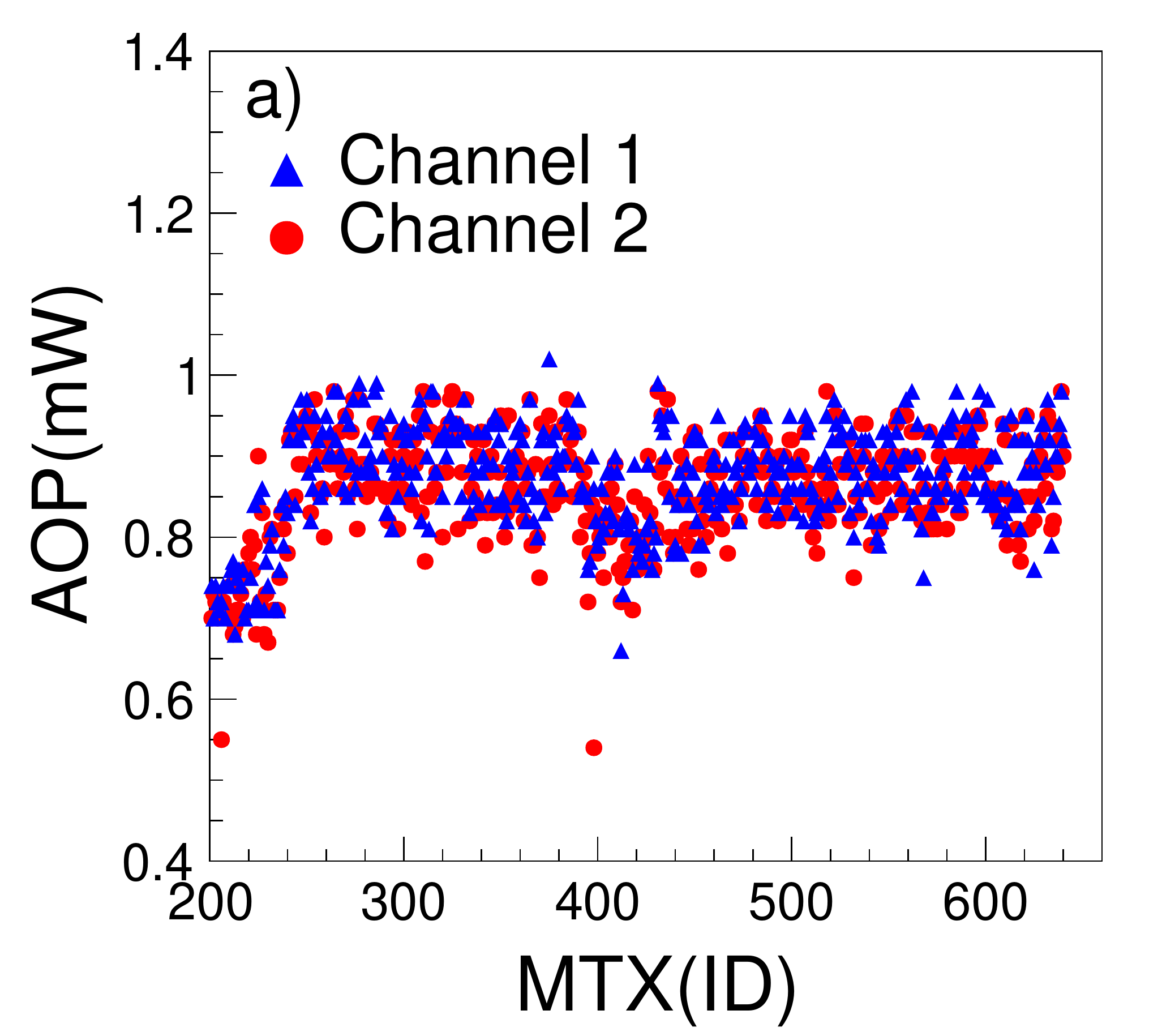}
  \includegraphics[width=.49\linewidth]{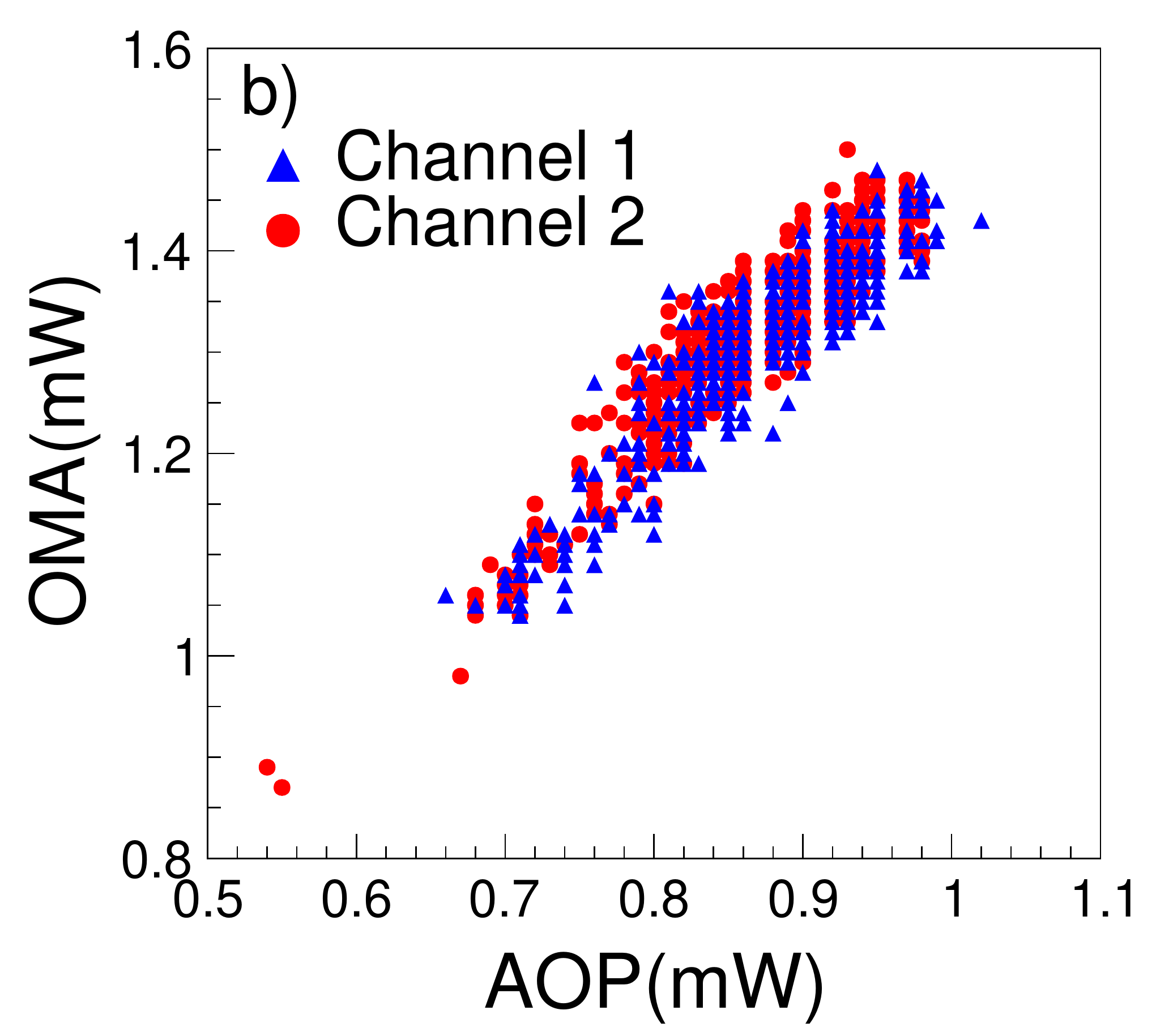}  

  \includegraphics[width=.49\linewidth]{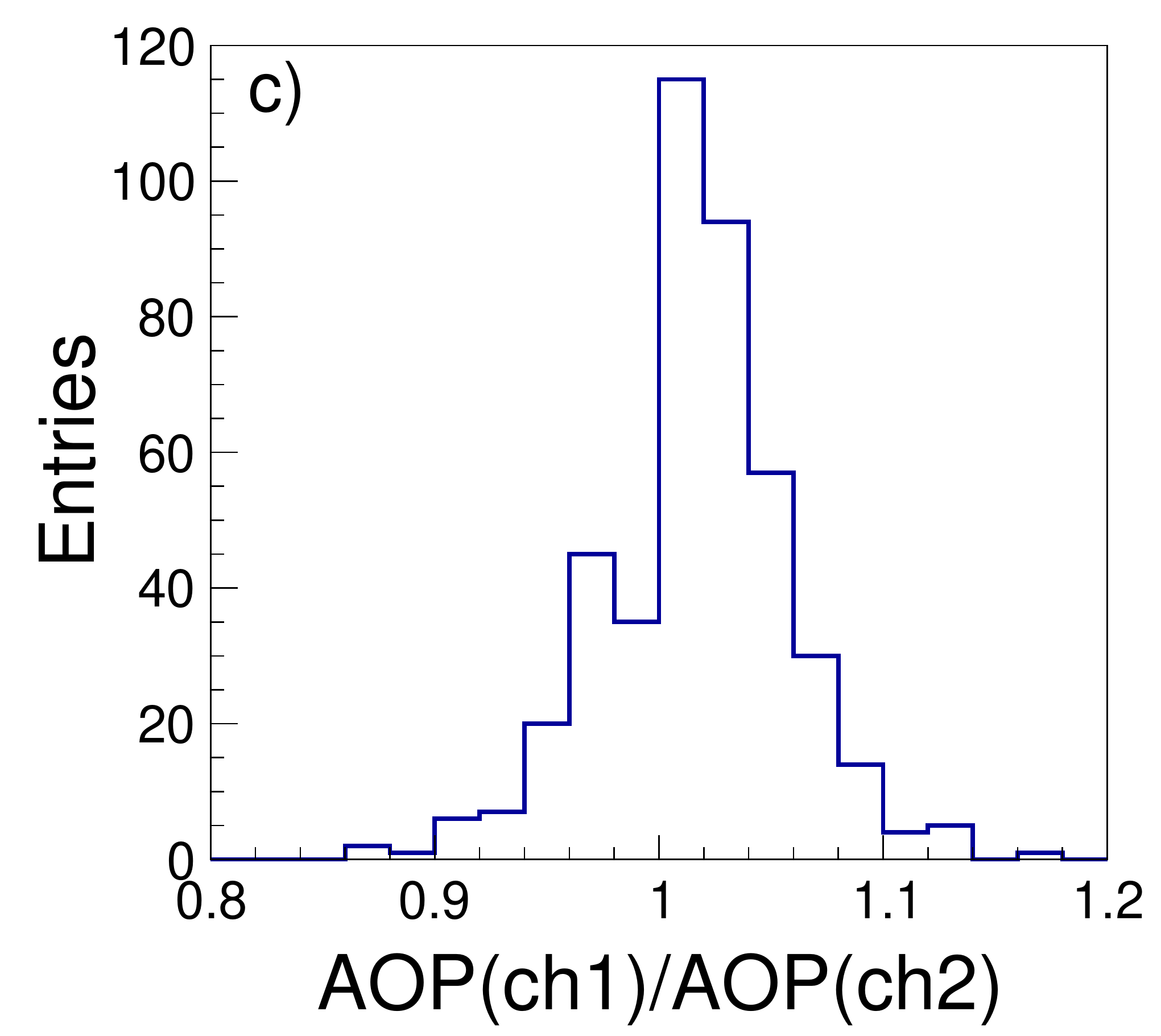}
  \includegraphics[width=.49\linewidth]{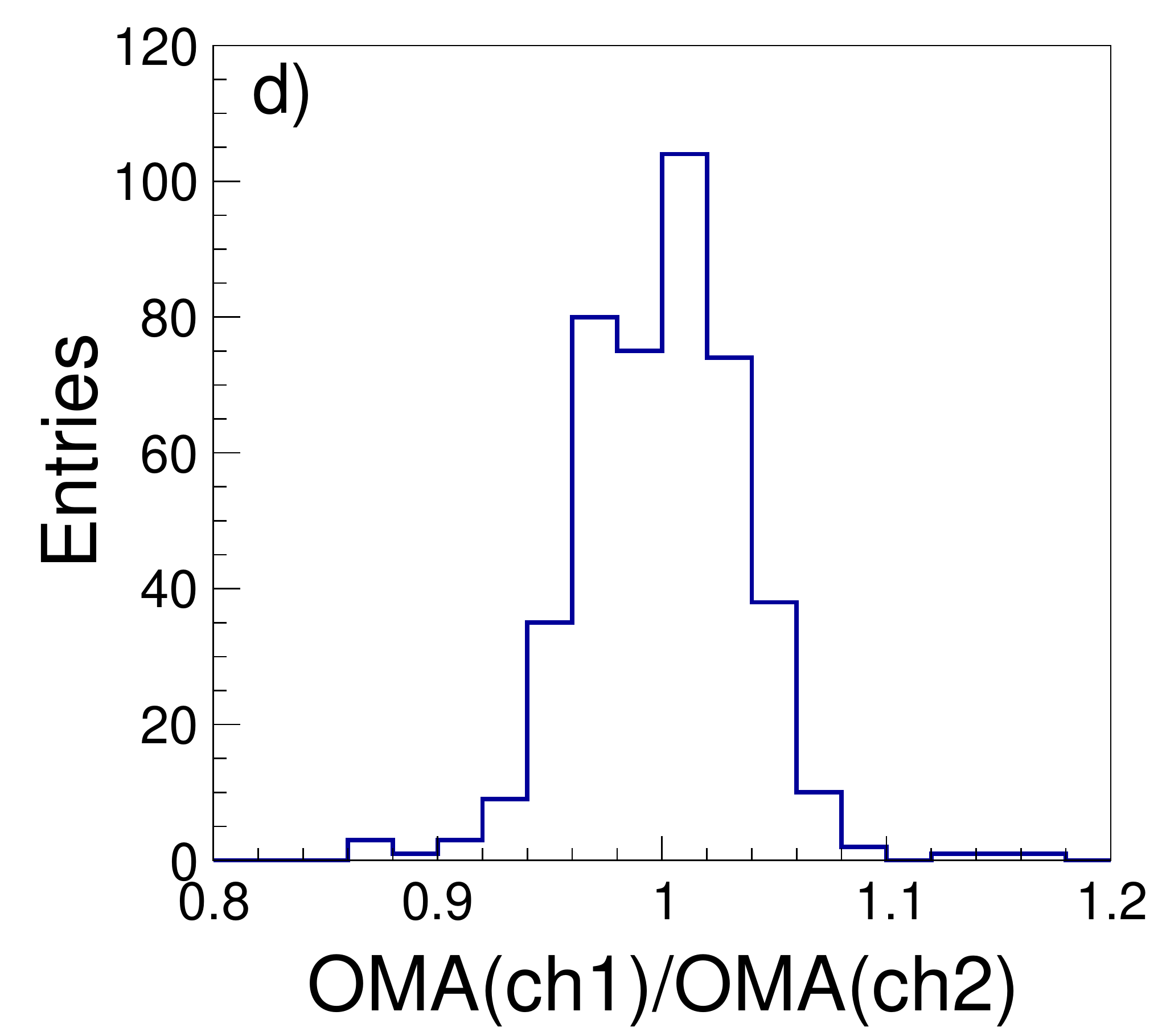}   
  
  \caption{ {\color{black} A pioneer batch of 440 MTx modules 
  were tested at 8.5~Gbps. 
  The distributions are shown for the eye-diagram parameters of
  a)  AOPs in sequence of module number,
  and b) correlation of OMA versus AOP.}
  For each MTx with two transmitter channels,
  the ratio of the {\color{black} AOPs and OMAs} are plotted in c) and d), respectively.
  {\color{black} The standard deviation of each is 4~\%.}
  \label{fig:MTx_AOPOMA} }
\end{figure}

\begin{figure}[b!] 
  \centering
  \includegraphics[width=.49\linewidth]{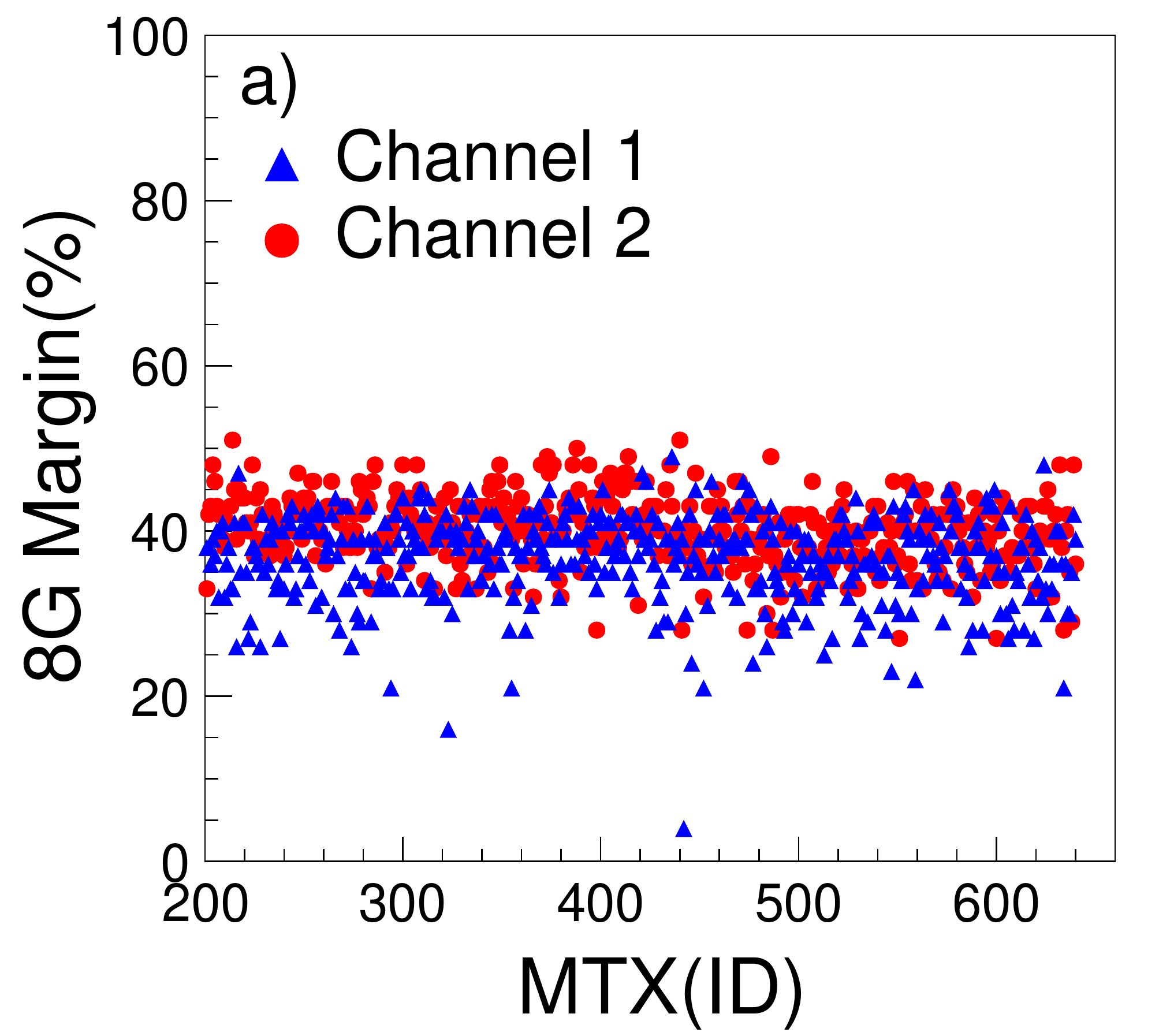}
  \includegraphics[width=.49\linewidth]{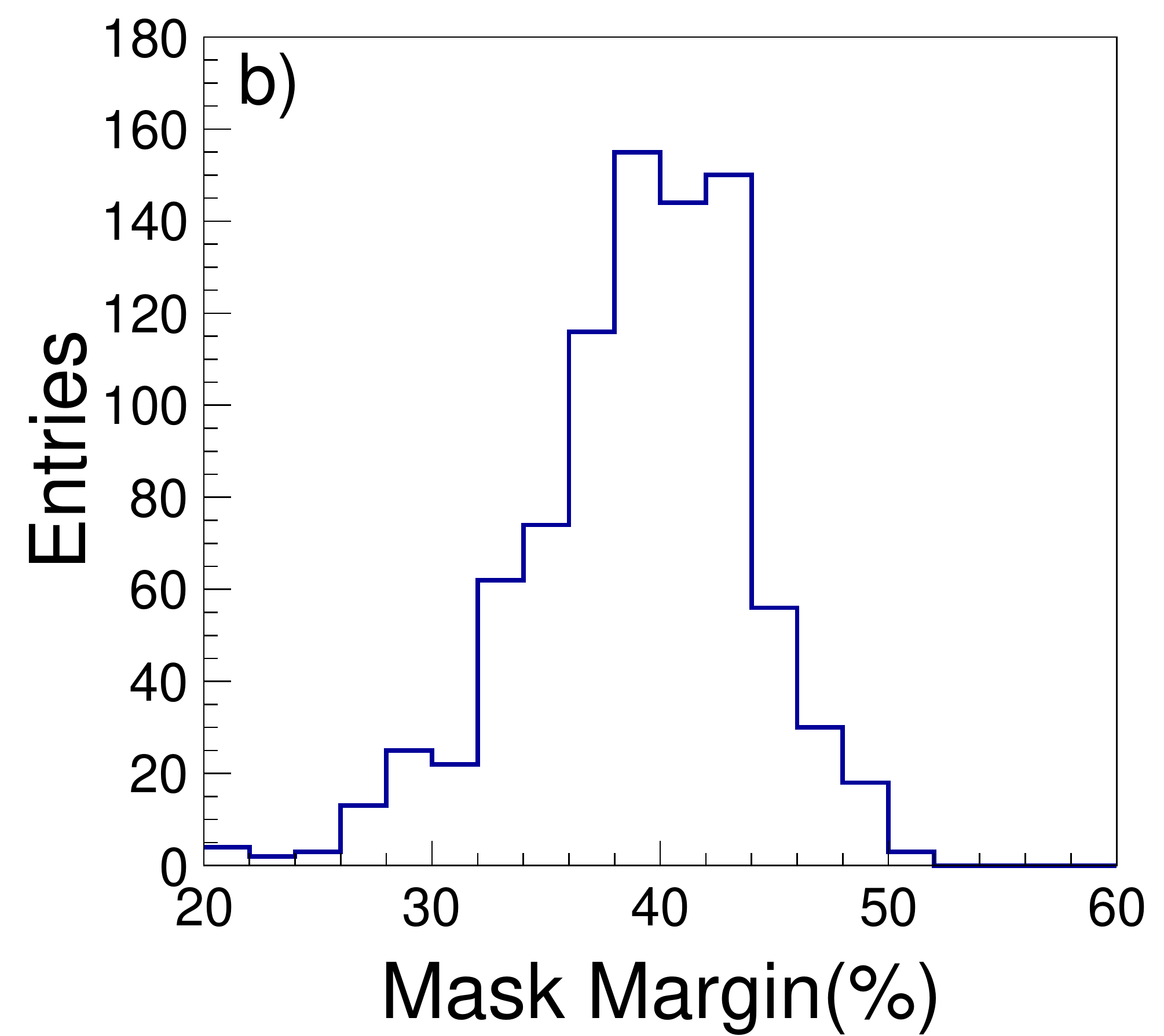}  

  \vspace{.0cm}
  \caption{ 
  The eye-diagrams of the pioneer batch of 440 MTx modules
  are plotted for a) the mask margins at 8.5~Gbps versus the module number,
  and b) the distribution of mask margins with
  a mean of 39 \% and standard deviation of 5 \%.
  \label{fig:8G_margin} }
\end{figure}

{\color{black}
The transmitter outputs deviate mostly on the light power level. 
{\color{black} This is managed by sorting the TOSAs by light power 
for module assembly.}

The uniformity of transmitter outputs is investigated with 
a pioneer batch of 440 MTx modules. The test was conducted
at 8.5~Gbps with eye-diagrams measured
by a BER sampling oscilloscope (MP2100B, Anritsu).
The eye-diagram parameters including the AOP and OMA  are analyzed.
The distribution of AOP is plotted in 
Fig.~\ref{fig:MTx_AOPOMA}.a.}
The deviation in sequence of module number is associated with the 
TOSAs of different {\color{black} delivery batches being sorted by light power.}
The OMA shows a linear correlation to the optical power. 
The distribution of OMA versus AOP is shown in Fig.~\ref{fig:MTx_AOPOMA}.b.

{\color{black}
The two TOSAs assembled on a MTx module were chosen
with light powers being consistent within 3~\%.
Plotted in Fig.~\ref{fig:MTx_AOPOMA}.c and \ref{fig:MTx_AOPOMA}.d 
{\color{black}
are the distributions of ratios of AOPs and OMAs obtained from each MTx, respectively.}
The standard deviations of both plots are 4~\%. 
The wider distributions is caused by 
{\color{black} the systematics on light-coupling connecting 
fiber-optic cable to TOSAs.} 
The effect is {\color{black} calibrated with a light source
and power meter (CMA5, Anritsu)}.}
The systematic error on TOSA light power measurement is estimated
to be 6~\%. 

The bias currents of the LOCld to VCSELs may have contributed
to the widening distributions of the ratios of optical output parameters.
It is estimated to be less than 3~\% assuming the fluctuation is 
caused entirely by the LOCld output current.

The uniformity of transmitter channels is best observed with
the mask margin of eye-diagrams. 
{\color{black} The distribution is shown in Fig.~\ref{fig:8G_margin}.}
Despite the large deviation on optical light powers, the  
{\color{black} mask margins measured at 8.5~Gbps}
are uniformly distributed  
{\color{black} around 39~\% 
with a standard deviation of 5~\%.}

\begin{figure}[b!] 
  \vspace{.2cm}
  \centering
  \includegraphics[width=.8\linewidth]{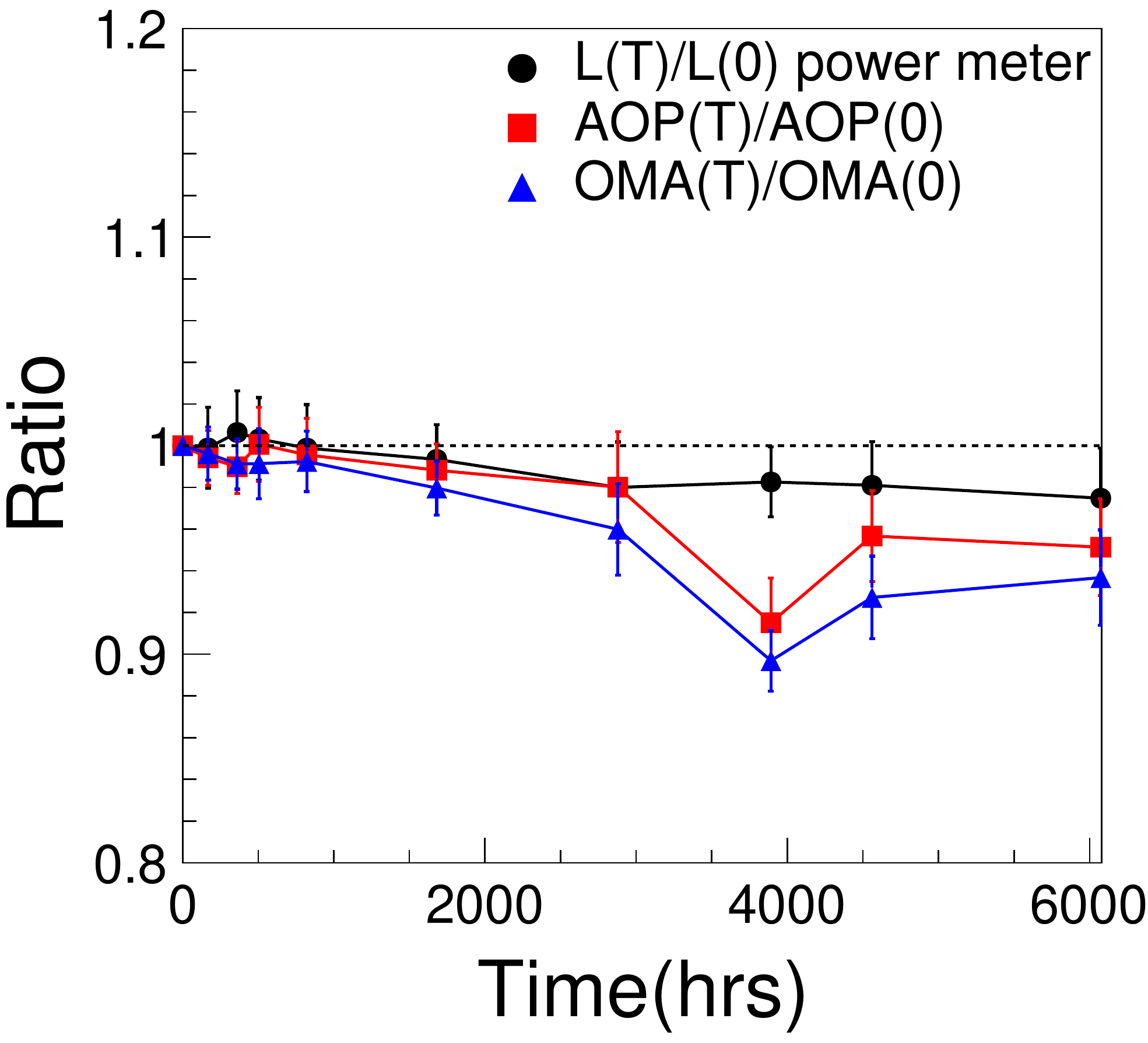}
 
  \vspace{-.3cm}
  \caption{ The burn-in of 24 MTx modules are measured periodically
  for the eye-diagrams at 10~Gbps.
  The optical outputs are also measured by a power meter. 
  {\color{black} The ageing effect is monitored for the AOP and OMA of eye-diagrams 
  and the {\color{black} power} meter readings, relative to the initial values.
  Plotted are the means with RMS errors of the 48 transmitter outputs.}
  The dips of AOP and OMA near 4k hrs are caused by 
  {\color{black} a fatigue compressing clip of a fiber connector in use.}
  \label{fig:ageing} }
\end{figure}

\section{Ageing test} 
\label{sec:ageing}

The ageing effect in the MTx is monitored with burn-in of a small batch of 
24 modules in room condition.
{\color{black}
These modules are powered on continuously with the 
transmitter light power levels 
measured by optical power meters. 
The purpose is to detect early indication of degradation, which 
does not require a large statistics of modules.

The test modules were examined periodically for 
bit-error-rate and eye-diagrams {\color{black} at 10~Gbps}.
Each of the modules was dismounted briefly from the burn-in setup
to be connected with differential inputs and fiber cable to oscilloscope.}
 
The ageing of VCSELs {\color{black} is expected for}
light power deviation to higher or lower level. 
{\color{black}
The ageing of LOCld is monitored for bit-error rate and eye-diagrams.}
The burn-in has accumulated for over 6k hours with no error observed.
Plotted in Fig.~\ref{fig:ageing} are the measurements in time 
for the average optical powers and the modulation amplitudes 
of the eye-diagrams relative to the initial values. 

{\color{black}
The eye-diagram measurement of the transmitter outputs 
has a large uncertainty 
due to light coupling with fiber-optic cable.} 
As a cross-check, the transmitter outputs are also measured 
with a {\color{black} optical} power meter using a different fiber-optic cable.
In one occasion, the light powers of eye-diagrams were  
significantly lower (Fig.~\ref{fig:ageing}, near 4k hrs).
It was found due to {\color{black} mechanical fatigue of a}
compressing clip on a fiber connector.

Over the burn-in of 6k hours, the light power distribution shows a slight 
degradation by less than 5 \%. 
This is considered mostly due to the VCSEL intrinsics over time.
The eye-diagrams observed show no indication of deterioration
{\color{black} in data transmission.}

\section{Summary} 
\label{sec:sum}

The MTx and MTRx optical transceivers are developed for the 
ATLAS Phase-I applications. 
The driver ASIC and opto-electronics are evaluated for durability.
The production process had the assurance {\color{black} checking} on components and
the quality control on modules to improve module reliability and 
uniformity.
The ageing effect is monitored for more than 6k hours in burn-in.
The optical outputs show a genuine light power degradation within 5 \%,
without bit-error in data transmission.      


{}

\end{document}